\documentclass[12pt,letterpaper]{article}
\usepackage{amsmath,amssymb,pgf,pgfarrows,pgfnodes,float,appendix, hyperref}
\usepackage{graphicx}
\usepackage{subfigure}
\usepackage[margin=0.9in]{geometry}

\newcommand{\beq}{\begin{equation}}
\newcommand{\eeq}{\end{equation}}

\newcommand{\be}{\begin{equation}}
\newcommand{\ee}{\end{equation}}
\newcommand{\bea}{\begin{eqnarray}}
\newcommand{\eea}{\end{eqnarray}}

\title{{\rm\footnotesize \qquad \qquad \qquad \qquad \qquad \ \qquad \qquad \qquad \ \ \ \ \ \                  RUNHETC-2020-4, UTTG-01-20 }\vskip.5in    Holographic Space-time, Newton's Law, and the Dynamics of Horizons }
\author{Tom Banks\\
Department of Physics and NHETC\\
Rutgers University, Piscataway, NJ 08854\\
E-mail: \href{mailto:banks@physics.rutgers.edu}{banks@physics.rutgers.edu}
\\
\\
Willy Fischler\\
Department of Physics and Texas Cosmology Center\\
University of Texas, Austin, TX 78712\\
E-mail: \href{mailto:fischler@physics.utexas.edu}{fischler@physics.utexas.edu}}
\date{}
\begin{document}
\maketitle

\begin{abstract}
We revisit the construction of models of quantum gravity in $d$ dimensional Minkowski space in terms of random tensor models, and correct some mistakes in our previous treatment of the subject. We find a large class of models in which the large impact parameter scattering scales with energy and impact parameter like Newton's law.  
The scattering amplitudes in these models describe scattering of jets of particles, and also include amplitudes for the production of highly meta-stable states with all the parametric properties of black holes.
These models have emergent energy, momentum and angular conservation laws, despite being based on time dependent Hamiltonians.  The scattering amplitudes in which no intermediate black holes are produced have a time-ordered Feynman diagram space-time structure: local interaction vertices connected by propagation of free particles (really Sterman-Weinberg jets of particles). However, there are also amplitudes where jets collide to form large meta-stable objects, with all the scaling properties of black holes: energy, entropy and temperature, as well as the characteristic time scale for the decay of perturbations.  We generalize the conjecture of Sekino and Susskind, to claim that all of these models are fast scramblers.  The rationale for this claim is that the interactions are invariant under fuzzy subgroups of the group of volume preserving diffeomorphisms, so that they are highly non-local on the holographic screen. We review how this formalism resolves the Firewall Paradox.
\end{abstract}

\section{Introduction}

This paper is a replacement for hep-th:/1606.01267, which has been withdrawn.  We will study a class of discrete time dependent Hamiltonian systems, which couple together more degrees of freedom as time goes on.  That is, the Hamiltonian has the form
\beq H(t) = H(-t) = H_{in} (t) + H_{out} (t) , \eeq where $H_{in} (t)$ is a function of canonical fermion variables\footnote{It is probably easy to generalize our considerations to models in which the fermion carries an additional label, and satisfies a more general super-algebra.} labeled by an $n$-th rank anti-symmetric tensor $\psi_{a^1 \ldots a^n}$.  The indices run from $1$ to $t$, which labels the discrete time.  $H_{out} (t)$ depends on the components of a similar fermionic tensor, in which the indices run from $1$ to $T$, but some of them have at least one component whose index is $> t$.   The fermions in $H_{in}$ anticommute with those in $H_{out}$. We will eventually be interested in the $T$ goes to infinity limit.  Note that $H_{out} (t = \pm T)$ vanishes.

To construct a Hamiltonian, we introduce the matrix \beq M_i^j = \psi_i^A \psi^{\dagger\ j}_A, \eeq where $A$ runs over $n-1$ of the anti-symmetric indices, and we have raised those indices for notational simplicity.  The Hamiltonian is of the form 
\beq H_{in} (t) = P_0 + \frac{1}{t} {\rm Tr}\ P(M/t^{n - 1}) , \eeq where the coefficients of the polynomial $P$ are $t$ independent in the large $t$ limit.  $P_0$ is complicated because its form depends on the constrained subspaces of the Hilbert space, which we are about to describe.  

The amplitudes that we will estimate are defined by starting at $-T$ in a constrained subspace of the Hilbert space defined by
\beq \psi_i^{A(b)} | Scatt \rangle = 0 , \eeq where $1 \leq b \leq k$ and the multi indices $A(b)$ are restricted to run over only $n_b$ values.   $i$ runs from $1$ to $T$.  For example, when $k = 1$ and $n = 3$ we are constraining \beq \psi_i^{ab} = - \psi_i^{ba}\eeq with $a,b$ running between $1$ and $n_1$.  From the construction of the matrix $M_i^j$, one can see that this constraint makes it block diagonal
\beq M = \begin{pmatrix} M_{n_1} & 0 \cr 0 & M_{T - n_1} .\end{pmatrix} \eeq Similarly, if the constraints involve $b$ non-overlapping index ranges, of sizes $n_1 \ldots n_b$, then the matrix will have $b + 1$ blocks, of sizes $n_1 \ldots n_b, T - \sum n_i$, when acting on the constrained subspace.  
The single trace construction of the Hamiltonian implies that the fermions with all indices in one of the (non-overlapping) ranges of $n_b$ indices, become independent, non-interacting systems at $- T$.  This is the reason that we have called these Scattering states.  We will argue that the final state at $T$ satisfies a similar set of constraints.  The objects of interest in this model will be amplitudes to go from some past scattering state to some future scattering state.  We insist that the number of constraints be much smaller than the total number of fermions, so $\sum_b n_b^{n - 1} \ll T^{n - 1}$. 

We should note that a small number of the constraints play the role of severing the connection between the $k$ independent systems, while of order $\sum_b n_b^{n - 1} T$ constraints sever the small blocks of the matrix from the large block of size $T - \sum n_b$.  We will abuse language and call all the fermions making up one of the small blocks of the matrix, block variables, or simply blocks. We would now like to argue that the final state is constrained, and that the number of constraints scales at large $T$ like $\sum_b n_b^{n - 1} T$.  That is, in the limit $T \rightarrow \infty$, $\sum_b n_b^{n - 1} $ is an asymptotic conservation law.  It commutes with the Scattering operator.   The argument has two parts.

First of all, we claim that the Hamiltonian ${\rm Tr}\ P(M/N^{n - 1})$ has a non-trivial large $N$ limit with energies that are of order $N^n$ and energy differences of order $1$.  To see this, note that the leading order diagram (Fig. 1)
\begin{figure}[h!]
\begin{center}
  \includegraphics[width=12cm]{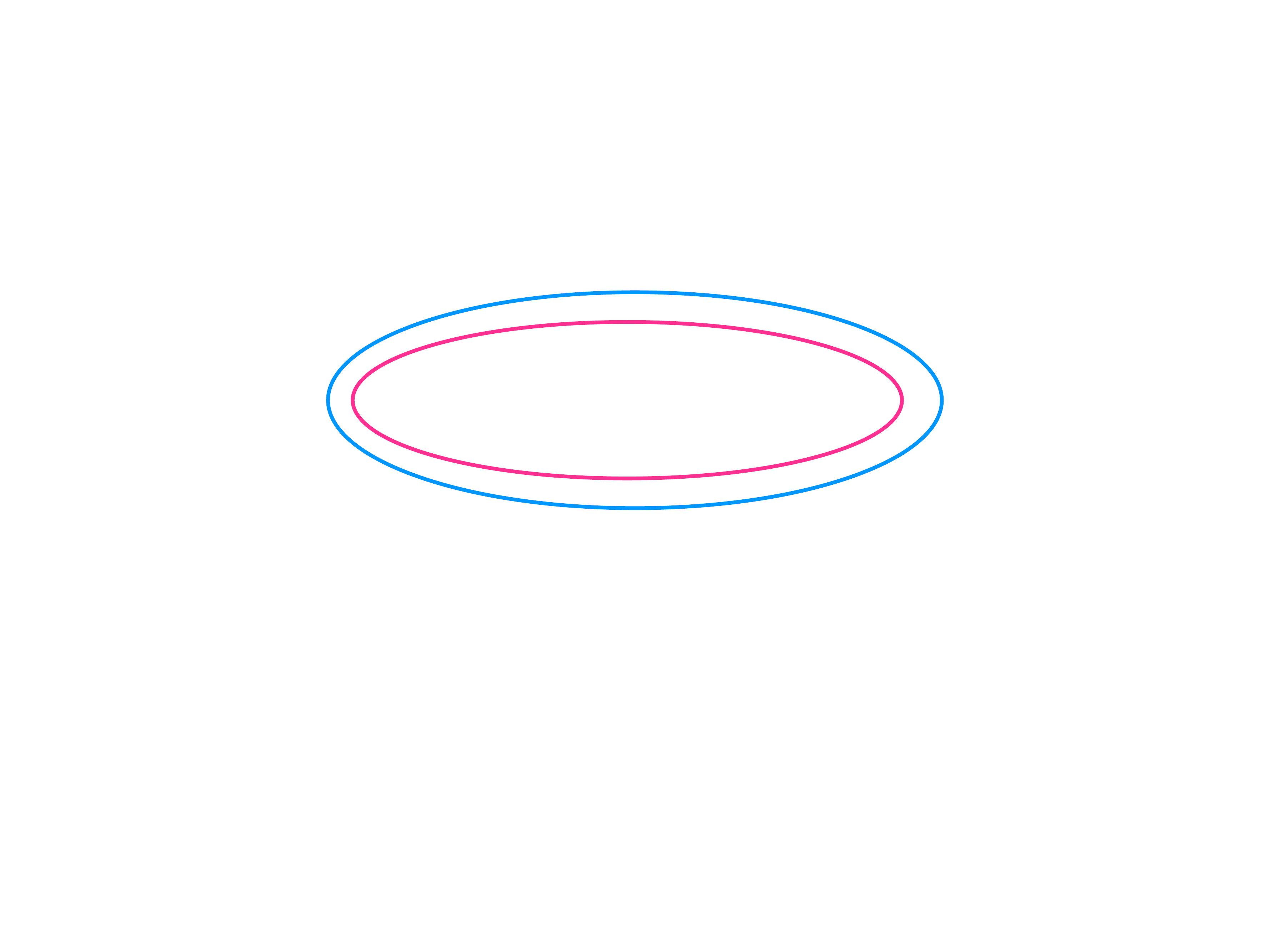}
\end{center}
\vspace{-1.6cm}
\caption{Leading order diagram.\label{fig1a}}
\end{figure}
  
for the free energy of this Hamiltonian scales like $KN$ where $K \sim N^{n - 1}$.  For large $N$ we can ignore the anti-symmetry requirement to leading order.  Higher order planar terms for $n = 2$ will all scale the same because this is just the 't Hooft limit of a matrix model.  For $n > 2$ the model is simpler.  Draw the fermion propagator as a double line with two colors (Fig. 2), 
\begin{figure}[h!]
\begin{center}
  \includegraphics[width=12cm]{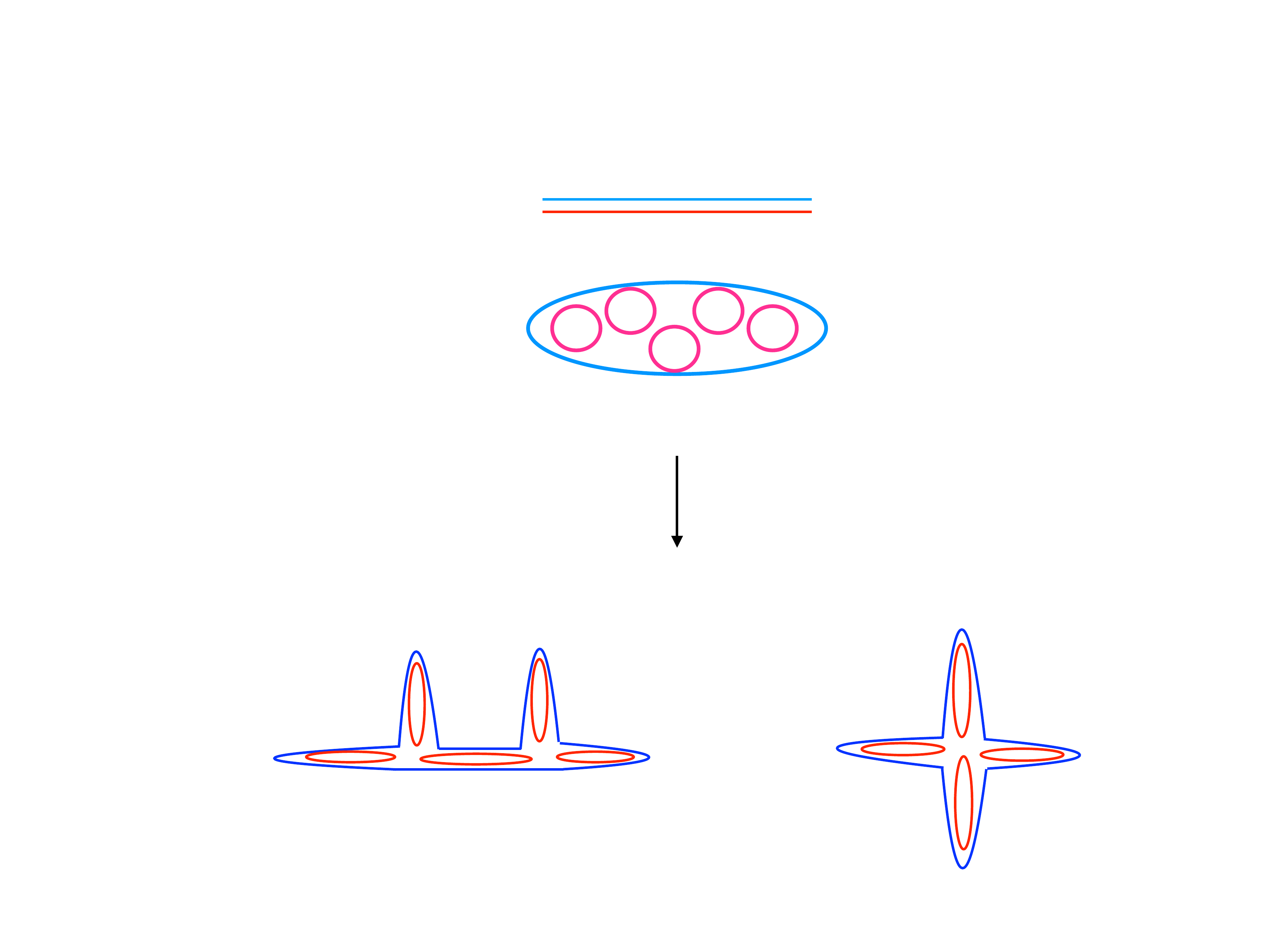}
\end{center}
\vspace{-1.6cm}
\caption{Higher order diagrams.\label{fig1b}}
\end{figure} 
the red line carrying $n - 1$ indices.  Then it is clear that the dominant scaling for $n > 2$ is like that of a vector model, namely amplitudes are dominated by cactus diagrams.  However, in double colored line notation it is more convenient to draw the graphs as a single blue line surrounding a collection of red loops.  These can be deformed into combinations of vertices with different numbers of fermions in a variety of ways.  If we write an interaction that is a function of the fermion bilinear divided by $K$ then all leading terms scale in the same way.  We will not study higher orders, which are complicated by the antisymmetry requirement.  Note however that the organization of this perturbation theory will differ from that of vector models.  Rather, they resemble rectangular matrix models with the small side scaling as a fractional power of the large side.  For $n = 2$ of course, the leading behavior is given by summing all planar diagrams,equivalently diagrams with any number of red and blue loops, pinched in all possible ways consistent with the interactions in the polynomial $P$.  Note that this counting is valid for the free energy at any temperature, which will be, in leading order, proportional to a fixed function of the temperature.  Thus the model contains many states with order $1$ energy differences in the large $N$ limit.

\begin{figure}[h!]
\begin{center}
  \includegraphics[width=12cm]{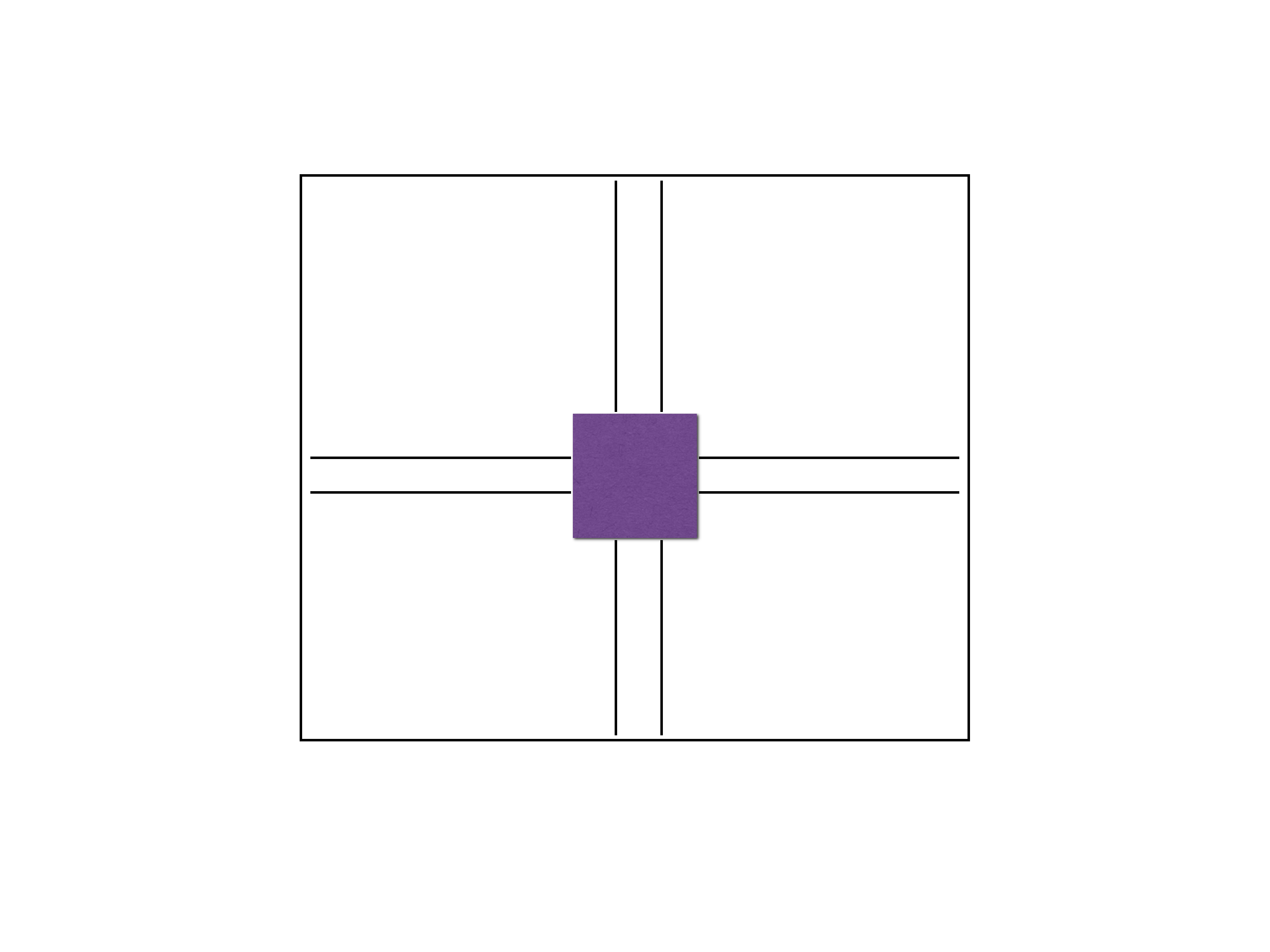}
\end{center}
\vspace{-1.6cm}
\caption{The matrix M obtained by gluing hypercubes together along a common boundary for $n = 2$.\label{squareinsquare2}}
\end{figure}

\begin{figure}[h!]
\begin{center}
  \includegraphics[width=12cm]{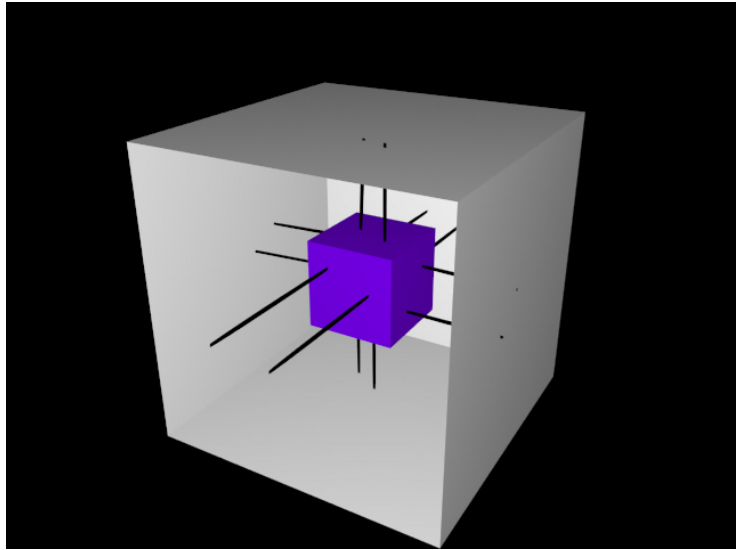}
\end{center}
\vspace{-1.6cm}
\caption{The matrix M obtained by gluing hypercubes together along a common boundary for $n = 2$, the front face has been made transparent to exhibit the gluing procedure.\label{cubeincube3}}
\end{figure}

There is a useful geometric interpretation of the rules of this class of large $N$ tensor models.   Think of each variable $\psi $ as a open subset of an $n$ - cube or sphere.  The matrices $M$ glue two such hypercubes together along a common boundary, as in Figures \ref{squareinsquare2} and \ref{cubeincube3} .

\begin{figure}[h!]
\begin{center}
 \includegraphics[width=12cm]{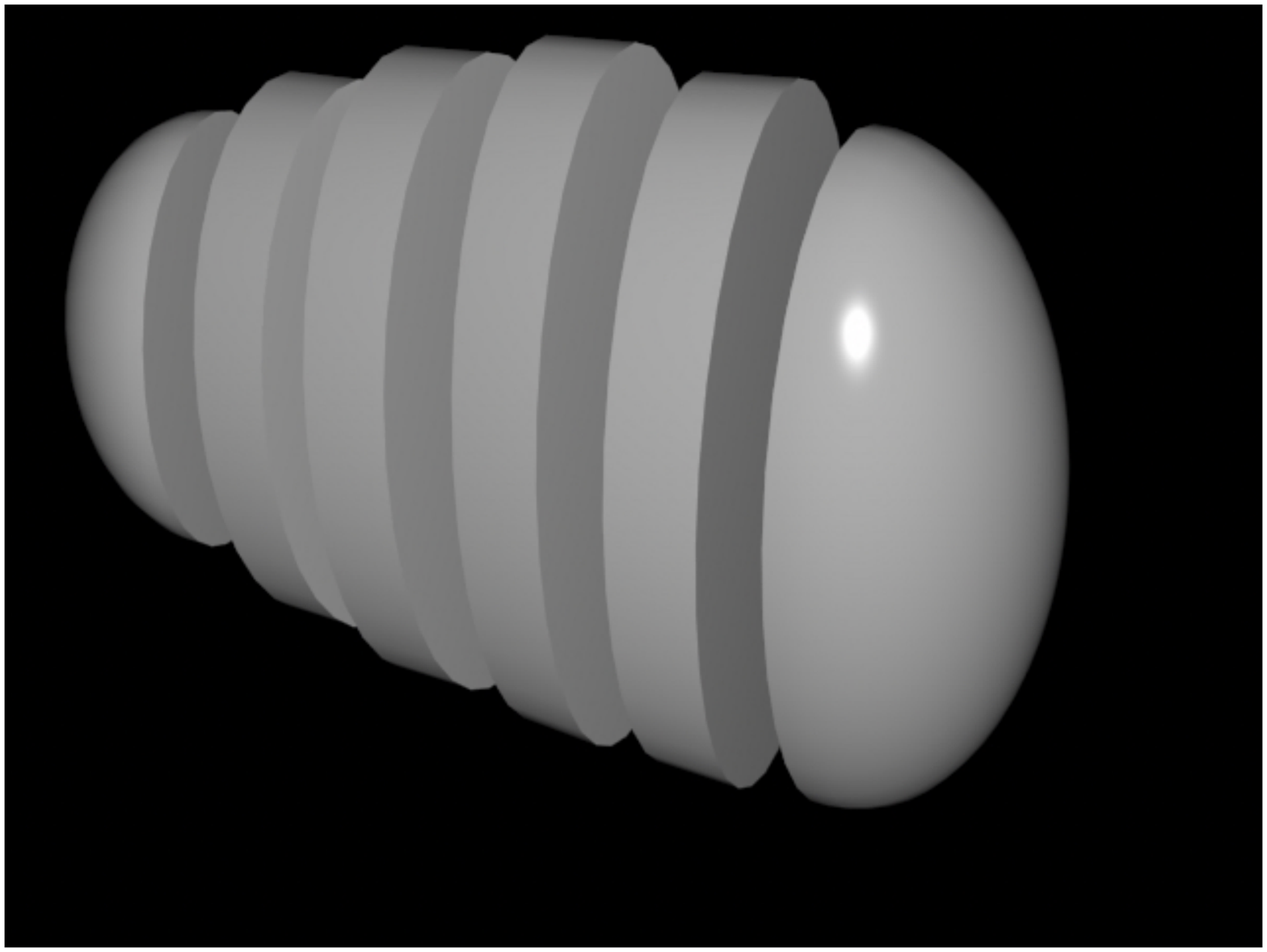}
\end{center}
\vspace{-1.4cm}
\caption{A monomial interaction. To simplify the picture, we did not attempt to illustrate the invariance under area preserving mappings, which could turn these regular slices into amoeba.\label{monomial}}
\end{figure}

We can think of a typical interaction in the Hamiltonian by opening up the trace, and thinking of this as picking a north and south pole on the $n$ sphere.  The first $\psi$ on the left is a patch near the north pole. Think of this patch as a fibration of $S^{n - 1}$ over an interval in polar angle.  This is glued to another ribbon of $S^{n - 1}$s and another, for the length of the polynomial.  The trace then eliminates the special choice of polar axis and the interaction is in fact invariant under the fuzzy version of the group of "area" (we use area as a shorthand for n-volume) preserving diffeomorphisms\footnote{Of course the word diffeomorphism is misleading.  Continuity and smoothness of functions have to do with the behavior of the large angular momentum Fourier coefficients and cannot be assessed in finite dimensional approximations.} on $S^{n}$.   A geometric picture of a monomial interaction is shown in Figure \ref{monomial}.
In particular, if we break the indices up into a group ($i$)whose number is $\ll N$, and the rest $A$, whose number is $o(N)$ then the variables $\psi_{1_1 \ldots i_{n - 1} A}$ represent the glue that connects a small $n$ cube to the rest of the volume. 

The reason that this picture is useful is that it illustrates the fact that our interaction is invariant under a fuzzy approximation to the group of area preserving mappings on the sphere\cite{tbjk}.  The $\psi$ variables are equal in number to the space of spinor sections on the sphere with a cutoff on angular momentum, which is equivalent to a cutoff of the Dirac operator.  Their commutation relations are invariant under $SO(n + 1)$, but also under a much larger group of unitary transformations on the indices.  Their bilinears can thus be thought of as fuzzy differential forms on the sphere, and the trace of products of bilinears is the integral of products of forms whose degree adds up to $n$.  This invariance property is the intuitive reason that all of these models are "fast scramblers"\cite{hpss}, since their interactions do not respect any notion of distance on the sphere.  More simply, one can see that every fermion is coupled to every other one by any trace interaction with four or more fermions.  

The conservation of what we will call Energy is a consequence of two aspects of our model.  First there is the fact that the interactions that remove or add constraints to the initial subspace all go to zero with time like $1/t$.  The polynomials have a fixed finite order, so our interactions are $2k-$local in the language of quantum information, with $k$ the highest power of $M_i^j$ that appears in $P$.  Thus it takes a time of order at least $t\ {\rm ln}\ t$ to remove $t$ constraints.   Moreover, during much of the interval $[-t,t]$ $H_{in} $ acts only on small subsets of the fermions.   We have not yet specified $H_{out}$ but we can insist that it act in a similar manner.  In the next section, we will argue that this follows from a natural consistency condition in the space-time interpretation of this quantum system.   

The fact that $ E = \sum_b n_b^{n - 1}$ is conserved says that at any time, the system has of order $ET$ constraints on the states in its Hilbert space.   Not all of these constraints will refer to variables acted on by $H_{in} (t)$ when $t$ is small.  This tells us that there must be constraints on the $H_{out} (t)$ Hilbert space.  We can now write the operator $P_0$.  Let $1 - \Pi$ be the projection on a particular constrained subspace of the Hilbert space, a particular subspace of scattering states with the same set of initial constraints.  Then, in that subspace, 

\begin{align*}
P_0 (t) &=  (1 - \Pi)[ \sum_{b^{in}} (n_{b^{in}}^{n - 1} + \frac{1}{n_{b^{in}}} {\rm Tr}\ P (t, M_{b^{in}\times b^{in}})) \\
  &\qquad +  \sum_{b^{out}} (n_{b^{out}}^{n - 1} + 
\frac{1}{n_{b^{out}}} {\rm Tr}\ P (t, M_{b^{out}\times b^{out}})) + \frac{1}{t} {\rm Tr}\ P (t, M_{t\times t}) ](1 - \Pi)
\end{align*}

Our remarks about energy conservation imply that if not all of the energy comes from blocks in $H_{in} (t) $ then $H_{out} (t)$ has to contain terms of the form

\beq \delta H_{out} (t) = (1 - \Pi) [\sum_{b^{out}} (n_{b^{out}}^{n - 1} + \frac{1}{n_{b^{out}}} {\rm Tr}\ P (t, M_{b^{out}\times b^{out}})  + \frac{1}{T} {\rm Tr}\ P (T, M_{T\times T}) ] (1 - \Pi) . \eeq
The phrase {\it has to} in the previous sentence is a bit of an exaggeration at this point, and will follow from our space-time consistency condition.  However, even from an abstract quantum mechanics point of view it is natural because when $t$ gets larger the $b^{out}$ blocks will be incorporated into $H_{in}$.  

\subsection{The leading large $t$ interaction between small blocks}

Now recall that the Hamiltonian of our model is the "'t Hooft" Hamiltonian multiplied by $1/t$.   This means that all of the 't Hooft couplings in the model are small for large $t$, so the model becomes weakly coupled.   Our goal is to write down the leading interaction between a pair of blocks of sizes $n_{1,2}$ coming from the Hamiltonian $H_{in} (t_*)$ when $1 \ll t_* \ll T$.  We will assume that for $s < t_*$ the constraints defining the blocks, and the block variables themselves, are not included in the Hamiltonian $H_{in} (s)$.  We'll also ignore the possiblity that there are other blocks in the system. Thus, the interactions between these two blocks can be computed by a sequence of computations of the type we do here, for $s \geq t_*$.  In the spacetime interpretation of the model, $t_*$ will be the impact parameter in the scattering amplitude.  

To begin the computation we write the time evolution operator for a single discrete time step as
\beq e^{ - i H (t_*)}  = \oint e^{ - i z} \frac{dz}{2\pi i (z - H(t_*))} , \eeq where the contour encircles the spectrum of $H(t_*)$ .  Define $H_0 = (1 - \Pi) H(t_*) (1 - \Pi) + \Pi H(t) \Pi $  and $H(t_*) = H_0 + V$.  $\Pi$ is the projector on the orthogonal complement of the particular constrained subspace in which exactly these two blocks begin at time $ - t_*$ with no interaction.   Every term in $V$ is of order $1/t_*$ or smaller and $V = 0 $ in the constrained subspace, as well as its orthogonal complement. Using the two by two block form of the operators, we can write the exact equation
\beq (1 - \Pi)(z - H(t))^{-1}(1 - \Pi) =[1 -  (z - H_0)^{-1}V \Pi (z -  H_0)^{-1}\Pi V]^{-1} (z - H_0)^{-1}   . \eeq  The operator $ V \Pi (z -  H_0)^{-1}\Pi V$ acts only in the constrained subspace. By construction, $\Pi (z - H_0)^{-1} \Pi$ contains no interaction between the two blocks under study.

The interaction between the blocks is mediated by perturbations that lift and restore the constraints.
Since the time interval is small in each computation in the above sequence, and the interaction is small, we can compute the amplitude by simply computing the effective Hamiltonian in the constrained subspace, due to the fact that part of the Hamiltonian, $V$,  proportional to $1/t_*$ does not commute with the constraints. The above computation of the matrix elements of the resolvent shows that the effective Hamiltonian in the constrained subspace is

\beq H_{eff} = (1 - \Pi) [H_0 + V \Pi (z - H_0)^{-1} \Pi V] (1 - \Pi) . \eeq The effective Hamiltonian is defined by 
\beq (1 - \Pi) (z - H)^{-1} (1 - \Pi) = (1 - \Pi) (z - H_{eff})^{-1} (1 - \Pi) ,  \eeq where $H_{eff}$ acts only in the constrained subspace.  $H_{eff}$ is $z$ dependent, reflecting the fact that the full Hamiltonian is not block diagonal.

$H_0$ has terms of order $1$ and terms of order $1/t_*$, while the second term in $H_{eff}$ is nominally of order $t_*^{-2}$.  However, this ignores small energy denominators, of order $1/t_*$.
 There are indeed such energy denominators since we can obtain a state in the orthogonal complement of the constrained subspace, by exciting one constrained variable of the form $\psi_i^A$ with $A$ being in one of the ranges corresponding to the small block variables and $i$ belonging to the large block. 
 
 To leading order the integral over the resolvent that gives the time evolution operator in the constrained subspace is a sum over the poles at the eigenvalues of $(1 - \Pi) H_0 (1 - \Pi)$. Near these poles, the second order effective Hamiltonian contains a factor $t_*$ from eigenvalues of $\Pi H_0 \Pi$ that are near those of $(1 - \Pi) H_0 (1 - \Pi)$. This factor cancels one of the factors of $1/t_*$ in $V$.  
 
 Now let's examine the dependence of the effective Hamiltonian on the variables in the small blocks.  $V$ itself is a sum of terms depending only separately on each of the blocks\footnote{We ask readers to beware of confusing blocks of the matrix, from the block diagonal form of operators in Hilbert space.} in the matrix.  This is simply a property of traces of powers of a matrix written in block form.  Thus in the second order effective Hamiltonian the terms coupling the two blocks come from terms in each $V$ that depend on different blocks.  At higher orders we can get more complicated combinations, but as we'll see, multibody interactions between blocks are suppressed by higher powers of $t_*^{- (n - 1)}$ than the leading two body term.  
 
From the equation for $H_{eff}$ we can make a remark about the sign of the operator.
 If the eigenvalues of $H_0$ in the orthocomplement of the constrained subspace are larger than those in the subspace itself (really one needs only weighted sums of eigenvalues to satisfy this inequality),  then the interaction operator is negative definite.
 We do not know at the present time whether this follows for a general Hamiltonian in our class, or represents a complicated inequality on the couplings.  
 
\begin{figure}[h!]
\begin{center}
  \includegraphics[width=12cm]{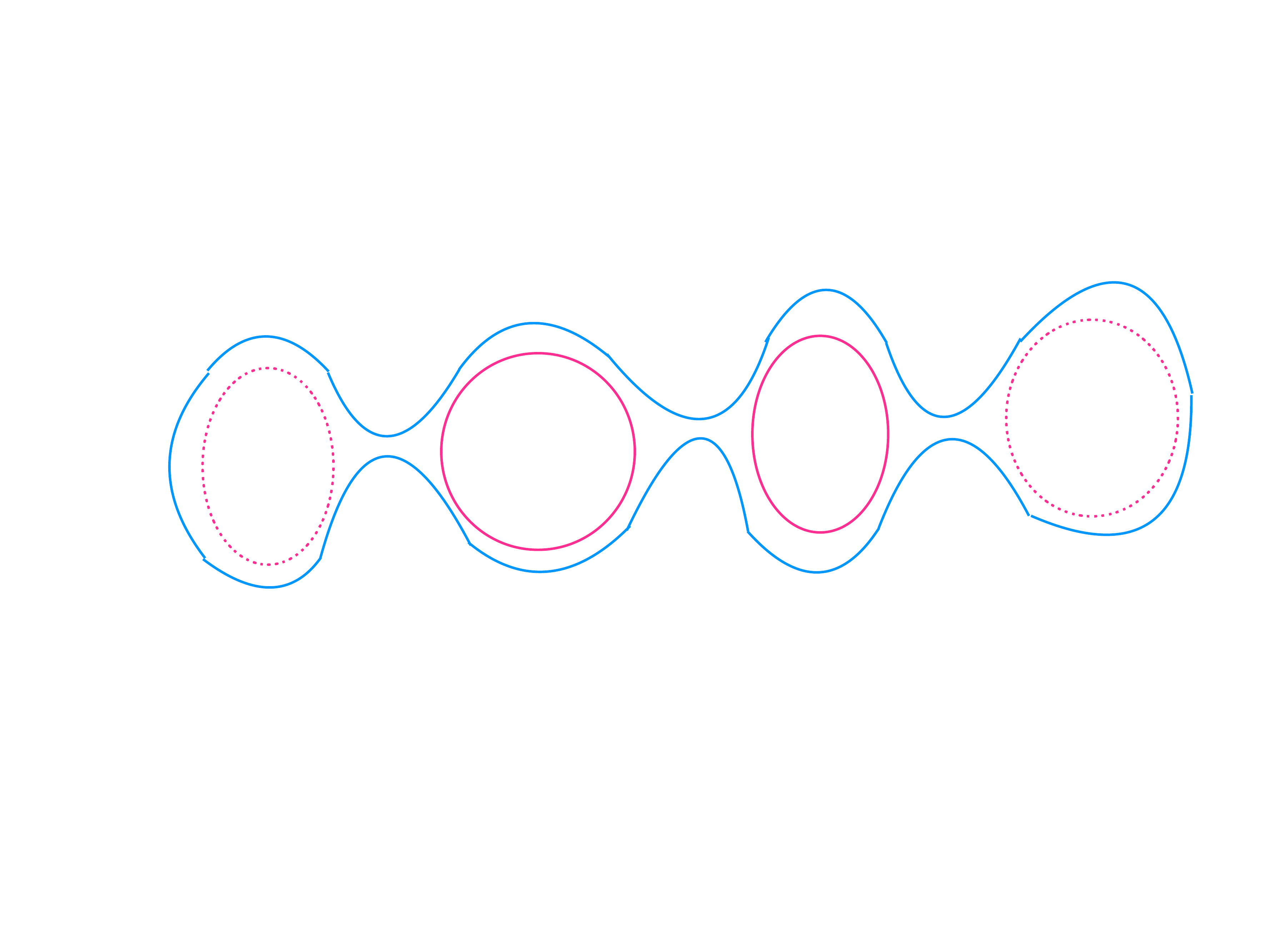}
\end{center}
\vspace{-1.4cm}
\caption{Implementing constraints.\label{constraints}}
\end{figure}

 We will need one more result to understand the scaling of the interaction with the block sizes and $t_*$.  This is simply the large $t$ scaling of the projection operator $(1 - P)$.  In path integral formalism, we can think of this as setting boundary conditions on the Grassman integration variables at the two ends of a time interval.  The states are functions on the fixed time Grassmann algebra and the constraint is 
 \beq \prod_{1,A} \delta (\psi_i^A (\tau) )\delta (\psi_i^A (0) )\delta (\bar\psi_i^A (\tau) )\delta (\psi_i^A (\tau) ) . \eeq We can implement the delta functions by integrating over auxiliary Grassmann variables $\eta_i^A (\tau, 0)$ and $\bar{\eta}_i^A (\tau, 0)$ and do the Gaussian integral with any quadratic term that couples only fermions with the same indices.  The result is given by a Feynman diagram like that of Fig. \ref{constraints} 
 The dotted red line counts the number of $A$ indices, while the solid blue line counts the number of $i$ indices.  In our problem, the $A$ indices are anti-symmetric tensors with indices that run from $1$ to $n_1$ and $1$ to $n_2$, while the $i$ index runs from $1$ to something of order $t$.  Thus the leading large $n_i$ and large $t_*$ scaling of the projectors on the constrained subspace with two blocks obeying the bound $n_1^{n - 1} + n_2^{n - 1} \ll t_*^{n - 1}$ is \beq 
 n_1^{n - 1} n_2^{n - 1} t_*^2  . \eeq  Thus, when comparing an amplitude with these two projection operators to a completely unconstrained calculation, we get a relative factor of \beq  n_1^{n - 1} n_2^{n - 1} t_*^{- 2 (n - 1)} \eeq
 
\begin{figure}[h!]
\begin{center}
  \includegraphics[width=12cm]{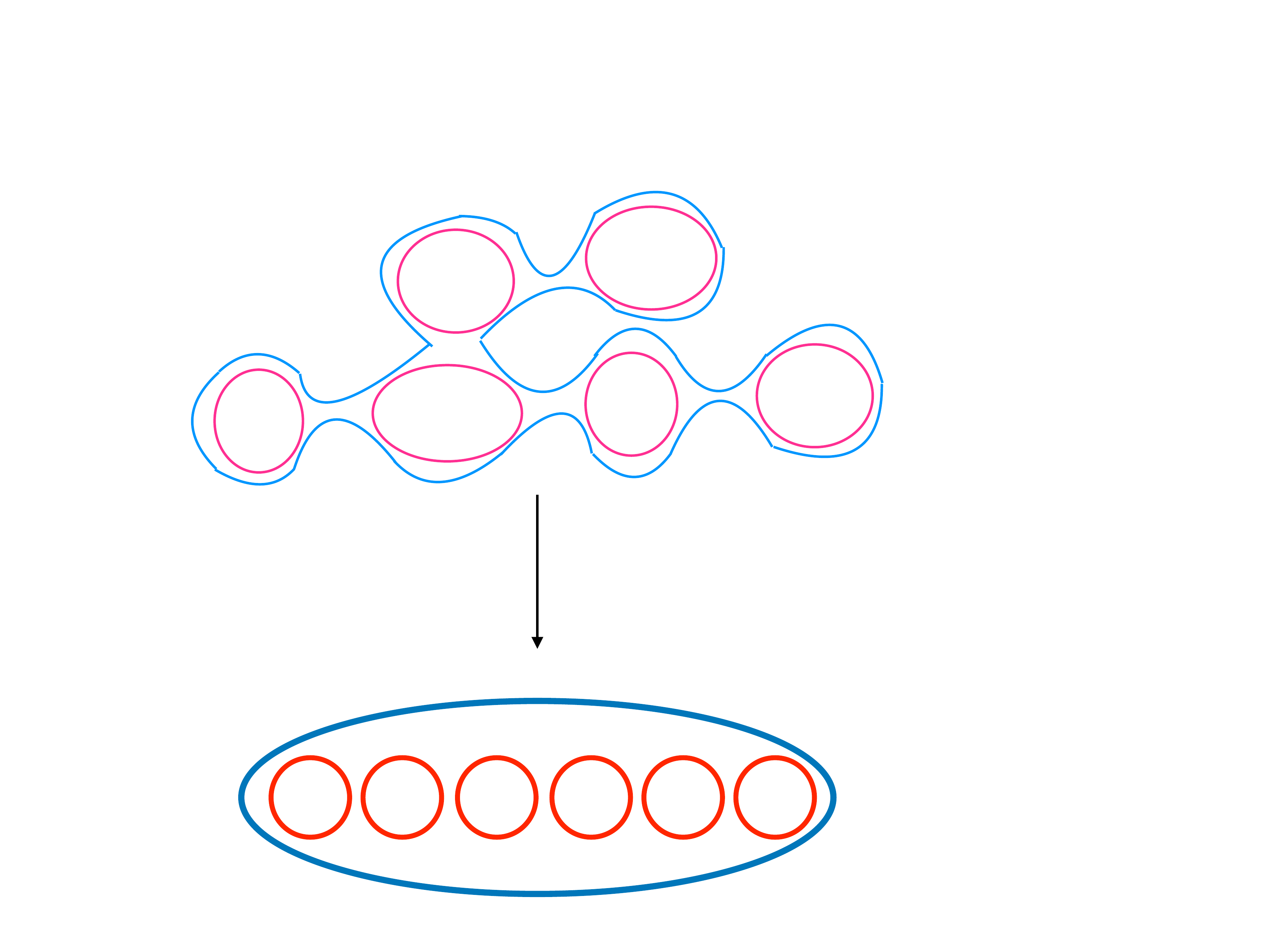}
\end{center}
\vspace{-1.4cm}
\caption{A prototype cactus diagram.\label{cacti}}
\end{figure}

 The interaction now scales like
 \beq  n_1^{n - 1} n_2^{n - 1}  \times t \times t_*^{-2(n - 1)} \times t_*^{-2} \times t_*^n . \eeq
 The last factor is the general large $t_*$ scaling of a free energy (a typical energy) in our large $t_*$ tensor models with 't Hooft couplings of order $1$.  The penultimate factor reflects the fact that we have scaled the interaction by a factor of $1/t_*$ relative to that large $t_*$ tensor model.   The next factor $t_*$ is the inverse of the small energy denominator, and finally we have suppression relative to a typical large $t_*$ diagram coming from the two projection operators.  The result is
  \beq  n_1^{n - 1} n_2^{n - 1}  \times t_*^{-(n - 1)}  . \eeq
 The scaling of a general diagram contributing to the interaction energy can be read off a simplified Feynman diagram like that of Fig. \ref{cacti} 
  In this diagram we do not show the explicit vertices, but simply the pattern of index loops.  Each of these diagrams can correspond to many different Feynman diagrams, depending on how we indent the blue line to form vertices.  The first diagram can only produce cacti, while if we allow any number of interior blue circles we can get arbitrary planar diagrams.  The dotted red circles each represent a small block and give a suppression factor $(\frac{n_b}{t_*})^{n - 1}$.   We see that at large $t_*$ multiblock interactions are suppressed.  They would correspond to replacing more solid red circles by dotted ones.
 
 To summarize, we've introduced a large number of tensor models which have a naturally defined scattering theory, because they have constrained states that decouple small subsets of degrees of freedom in the limit $ t \rightarrow \pm \infty$.  This is a consequence of a time dependent Hamiltonian, which couples together more degrees of freedom, organized as rank $n$ tensors with indices that run from $1$ to $t$, as the interval $[ - t, t ]$ gets larger.  Energy differences in the fixed time Hamiltonian scale like $1/t$, apart from the term  $P_0$, which is a sum over the small blocks.   The Hamiltonian itself is obviously not conserved, but we showed that there is an asymptotic conservation law:  if $C$ is the number of constrained q-bits, the limits $T\rightarrow \pm\infty$ of $C/|T|$ are equal to each other.
 
 The variables we use have a natural interpretation as an angular momentum\footnote{Equivalently a Dirac eigenvalue cutoff.} cutoff of the spinor bundle on the $n$ sphere\cite{tbjk} .  Using this language we can see that the fixed time Hamiltonians are invariant under the fuzzification of the group of volume preserving maps on the sphere, and are fast scramblers.  
 
 \section{Space-time interpretation of the models}
 
 In the previous section we studied, using purely quantum mechanical language, a class of finite quantum models, which have a scattering theory despite having no manifest spatial dimensions.  At $T \gg \sum E_i \rightarrow \pm \infty$, we exhibited a breakup of the Hilbert space into constrained subspaces, in each of which the asymptotic dynamics consists of a collection of non-interacting subsystems.  The largest of these, the horizon subsystem, becomes topological in the limit. All eigenvalues of this subsystem go to zero. The system has an asymptotically conserved quantum number, $\sum E_i$, which we called energy, and the horizon carries zero energy.  More importantly, in the asymptotic limit, the time dependent Hamiltonian of the decoupled horizon variables goes to zero.  
 
 In the limit, the variables of the theory approach generalized sections of the spinor bundle on the n - sphere.  They are also functions of the discrete positive variables $E_i$.  If we take the limit with fixed ratios of the $E_i$ then it is plausible that by tuning parameters in the Hamiltonian, the amplitudes become functions only of ratios of the $E_i$ and have finite limits.  It is plausible that one can tune the amplitudes to be conformally invariant on the sphere, or equivalently, Lorentz invariant on the light cone.
 The variables are not however quantum fields on the light cone in the Wightman sense.  They are not differentiable, since the zero momentum part had dynamics that was invariant under fuzzy volume preserving transformations on the sphere, before taking the limit.  We have speculated that they become Lorentz spinor operator half-measures on the sphere.  Bilinears in the spinors become measures, which transform as differential forms on the sphere.  
 
 Thus, despite the absence of space-time coordinates in the formulation of the theory, there is a natural interpretation of these models as a theory in space-time.  The nested tensor factors depending on time intervals in the quantum theory are identified with the Hilbert spaces of causal diamonds along a particular timelike geodesic in Minkowski spacetime.  
 The growing $n$ spheres in the quantum model are identified with the $n = d - 2$ dimensional holographic screens of causal diamonds along timelike geodesics in Minkowski space\footnote{It is clear from this sentence that space-time geometry is not a fluctuating quantum variable in this interpretation.  Many people have asked us about the fact that the geometry we assume does not seem to respond to the matter that is in it.  The answer to this question lies in Jacobson's hydrodynamic view of the origin of Einstein's equations and the space-time metric.  In a complex system, some parts of any particular process can be treated hydrodynamically, while we may need a more detailed description of more microscopic parts of the system.  In the example of black hole production, followed by Hawking radiation or the fall of a small system of "elementary particles" onto the black hole, we treat the whole system microscopically, as a scattering event in Minkowski space, up to the advent of black hole formation.  We then switch to a hydrodynamic description of the complex dynamics of the black hole while retaining the microscopic description of "particles" emitted from or absorbed by it.  In this paper, where we are not doing detailed calculations, we retain the microscopic description of the whole system, and treat the events as occurring in Minkowski space.  The physics is the same: the collision of "particles" to form a high entropy meta-stable state with which other particles can interact and from which they can be emitted/absorbed.}
 
 Here is a list of the properties of generic models from section 1, and their space time interpretation.
 
 \begin{itemize} 
 
 \item The models have a built in notion of causality.  Variables associated with a $d - 2$ sphere are isolated from the rest and have a number that grows like $t^{d-2}$.  Invoking the Covariant Entropy Principle\cite{bhgthjfsb} we identify the spatial radius of that sphere as proportional to $t$ in Planck units. The fixed $t$ Hilbert space is identified with that of a causal diamond of a proper time interval $[- t, t]$ along a geodesic in Minkowski space.
 
 \item The model has an asymptotic $R \times SO(d - 1)$ symmetry, with the $SO(d - 1)$ being picked out of the fuzzy group of volume preserving maps by a combination of the nesting of spheres and the constraints of the model.  In the limit $T \rightarrow \infty$ there is a scattering operator and we have argued that it is plausible that this operator acts only on the variables that make up the small blocks "liberated" from the majority of DOF by the asymptotic constraints.  In the limit $T \gg \sum E_i \rightarrow \infty $ with fixed ratios $E_i / E_j$, the variables converge to operator valued half measures on the momentum null cone, which transform as a collection of spinor fields under Lorentz transformations.  The scattering operator conserves the $SO(d - 1)$ subgroup of $SO(1,d - 1)$ but conservation of the rest of the group requires fine tuning of parameters in the Hamiltonian.  The rotation invariant asymptotic quantum number is interpreted as the energy.  It is proportional to the limit of $N_C/T$ where $N_C$ is the number of constrained q-bits.
 
 \item Meta-stable bound states are formed when the inequality $\sum E_i <  t^{d -2}$ is saturated in a causal diamond of proper time interval $[- t, t]$. Here $\sum E_i$ is the amount of asymptotically conserved energy that enters the diamond.  The time of formation of the bound states is of order $2t {\rm ln}\ t$ .   The definition of energy in terms of constraints implies that the probability of finding this high entropy meta-stable state to have of order $Et$ constraints after it has come into equilibrium is $e^{ - Et}$, which is thermal with temperature $\propto t^{-1}$.  This is interpreted as emission of "Hawking radiation" from the meta-stable equilibrium state.  Conversely, suppose we have, in the past half of a causal diamond of proper time interval $[- (t + \Delta t), (t + \Delta t)]$, 
 a state of the larger diamond that is a approximately a tensor product\footnote{The tensor product prescription is just approximate because the two subsystems have interacted in diamonds prior to $ - t - \Delta t$.} of the Hilbert spaces of degrees of freedom in two blocks of the matrix, of size $M \sim t$ and $m \ll t$, with constraints liberating these blocks from each other and from the $\sim (t + \delta t)^2 - M(t + \delta t)t$ degrees of freedom left in that diamond.  This state will evolve in a time of order $t {\rm ln}t$ to one in which the degrees of freedom linking the $M$ and $m$ blocks are excited and the full state is in equilibrium with entropy $(M + m)^{\frac{d - 2}{d - 3}}$.  This explains both the unexpected (from a quantum field theory point of view) increase in entropy in the process and the fact that the small system remains relatively unaware of the larger one for times of order $t$.  This argument, which uses only the fast scrambling nature of the interaction and the factor of $1/t$ in the Hamiltonian that equilibrates the system, resolves the firewall paradox\cite{fw} of the quantum field theory approach to black hole physics.  It is equally applicable to newly formed and old black holes.  The black hole interior, in this account, is erased in a time of order $t{\rm ln}\ t$ but recreated anew each time a small system falls on the black hole.  Another feature of black hole geometry that is reproduced by this model is the shrinking of the holoscreen volume  of causal diamonds inside the horizon.  That is, the holoscreen volume of a diamond that starts at time $\tau$ after horizon crossing, is a monotonically decreasing function of $\tau$.  In our models this is reproduced by the fact that the dynamics of in-falling matter is the dynamics within a small block.  Once the small and large blocks begin to come into equilibrium, via the excitation of off diagonal degrees of freedom, the size of the Hilbert space available to describe the interaction of small localized excitations decreases.
 
 \item The nesting of causal diamonds, which is incorporated via the time dependence of the HST Hamiltonian, combines with the definition of jet states in terms of asymptotic constraints, to give an understanding of why jets of particles are bulk localized objects.  We can follow the constraints from the largest causal diamonds to smaller ones and then back out to $T \rightarrow\infty$, and this defines the trajectories of incoming and outgoing particles in an emergent bulk space-time.  The mathematical definitions are all done in the quantum model.  A certain asymptotically non-interacting block of energy $E_i$ is defined by a constraint on variables in the system at $-T$.  Using the freedom of the volume preserving group we can define the degrees of freedom inside the small block to be localized in a spherical cap surrounding some particular point on the sphere, and the constrained variables to be those in an annulus surrounding it.  The area of that annulus is $E T$.  We can do the same for all the other small, asymptotically non-interacting, blocks, localizing them at different points.  The constraint $\sum E_i \ll T^{d - 3}$ guarantees that the area between these spherical caps is much larger than that in the caps themselves.  In going to smaller diamonds, there will be amplitudes where $\sum^{\prime} E_i$ remains much smaller than $t^{d - 3}$, where the primed sum might run only over some of the initial blocks.  We say that the blocks in the primed sum "enter into the past boundary of the small causal diamond" and we line up their angular positions with the ones defined on the largest diamond.  Invariance under the special rotation group picked out by this nesting follows from the volume preserving invariance of all Hamiltonians.  Thus, the models themselves define a notion of trajectories of weakly interacting objects, localized in angle, through the bulk of space-time, even though there is no "bulk" in the definition of the theory.
 
 \item As noted in the previous section, when the primed sum above is not equal to the original sum, we conclude from the asymptotic conservation of $\sum E_i$, that the rest of the energy appears at time $t$ as a set of non-interacting subsystems of the "out" Hilbert space.  The space-time interpretation of the models now has an important role to play in determining the structure of $H_{out}$.  The space-time interpretation of the model of the previous section is that it is the proper time dynamics along a particular time-like geodesic in Minkowski space.   The spacetime interpretation implies that there should be an identical Hamiltonian for every time-like geodesic.  Each of these is an independent quantum system.  The different initial conditions are constrained by an infinite set of constraints on mutual quantum information.  If we choose two intervals $[ - t_1, t_1], [ - t_2, t_2]$ along two different trajectories, the space-time picture implies that the causal diamonds of these intervals have some overlap. There is a causal diamond with maximal volume holographic screen, usually unique, in the overlap region.  Quantum mechanically this corresponds to tensor factors of equal dimension in the Hilbert spaces of the two systems.  Each system will prescribe a density matrix for that overlap, and the entanglement spectra of those density matrices should be equal.  This can be generalized to any set of geodesics. 
 
 This infinite set of pairwise constraints can be used in three different ways.  First of all, when all $t_i$ are equal and the energy in the causal diamond is less than the total incoming energy, the remainder should be found in some collection of causal diamonds, possibly overlapping.  Second, if some process occurs in causal diamond $1$ and diamond $2$ has no overlap with $1$, then the $H_{out}$ of diamond $2$ must describe the identical process.  Finally, we can generalize the constraints to intervals that are centered around different space-like hyperplanes in Minkowski space, which can give rise to amplitudes like that shown in Fig.\ref{exchange}
   A jet emitted from some past causal diamond can propagate to be part of the initial constraints on a different causal diamond in the future.  Thus, the model has amplitudes satisfying the clustering properties we usually derive from (time ordered) Feynman diagrams in quantum field theory.  We begin to see how field theory emerges as an approximate description of those amplitudes which do not lead to black hole production.
  
 The introduction of multiple versions of the dynamics corresponding to different time-like geodesics also leads to a derivation of space- translation and Lorentz boosts as asymptotic symmetries of the dynamics.  Unfortunately, while we can argue that space translation invariance will be satisfied for the large class of models defined in the previous section, the imposition of boost invariance imposes constraints that we have not been able to solve.  We also have evidence that boost invariance will not be satisfied for a generic choice of the polynomials in the previous section.  
 
 Consider the causal diamonds of proper time $t$ along two geodesics related by a Poincare transformation.  Let's choose the origin of proper time along the interval $[ - t , t]$ to be the same.  As $t \rightarrow \infty$ the overlap between the two causal diamonds is parametrically smaller than the areas of the individual diamonds.  For a generic state, Page's theorem\cite{page} then tells us that the state on the overlap is maximally uncertain.  However, we are not dealing with generic states.  Asymptotic energy conservation tells us that the asymptotic numbers of constraints are the same and that the number of constraints is much smaller than the total number of fermions.  We are then free to define the asymptotically non-interacting degrees of freedom in one system to be sitting at the Poincare transformed points on the sphere at infinity.   In order to do this and get a invariant result, we must of course take the $E_i$ to infinity at fixed ratio, with $T^{d - 3} \gg \sum E_i  $.  The definition of the Poincare transform of $E_i$ assumes that the jets are all massless.  The treatment of massive particles will be more complicated\footnote{From the evidence provided by string theory, we expect that all stable massive particles will either be BPS, or created in the collision of massive BPS particles and anti-particles, and have quantum numbers determined by a finite K-theory group.  The masses of the BPS particles are usually determined from the anti-commutators of the left and right supercharges.   We conjecture that the proper way to find the "K-theory" states is simply to explore the BPS particle anti-particle scattering matrix.  That is, we would find a violation of unitarity if we assumed there were no such states, and their masses will be determined by imposing unitarity.} .  Thus, the quantum information in the constrained subspaces, apart from information about the asymptotically decoupled and topological zero momentum modes is in subspaces of the same dimension and is totally contained in the overlap diamond, because the action of both Hamiltonians on states of non-zero energy is, in the limit indicated above, the same up to a Lorentz boost of the energy.  Thus, asymptotically, there should be a unitary transformation relating the two scattering operators, for every Poincare transformation.
 
 The phrase {\it should be} is not the same as the word {\it is}.  For time translations, rotations and space translations, there {\it is} such an asymptotic unitary in all of the models we have defined.  This is much less clear for Lorentz boosts, and we will present evidence below that it is not true for the generic Hamiltonian.
 
 \item When translated into space-time language, the calculation of the large time scattering of two jets that we did in the previous section, shows  that there is a Newtonian interaction between two jets, starting from the impact parameter of their trajectories, assuming they are straight lines determined by the asymptotic initial conditions, and following them out to infinity in both time directions.  It is clear from the space-time point of view that this calculation is valid only in the eikonal approximation, but this is expected because it was motivated by the large time limit, when the trajectories are far from each other.  
 
 There is another interaction between the two jets, coming from the "exchange diagrams" of Fig. \ref{exchange}
 \begin{figure}[h!]
\begin{center}
  \includegraphics[width=12cm]{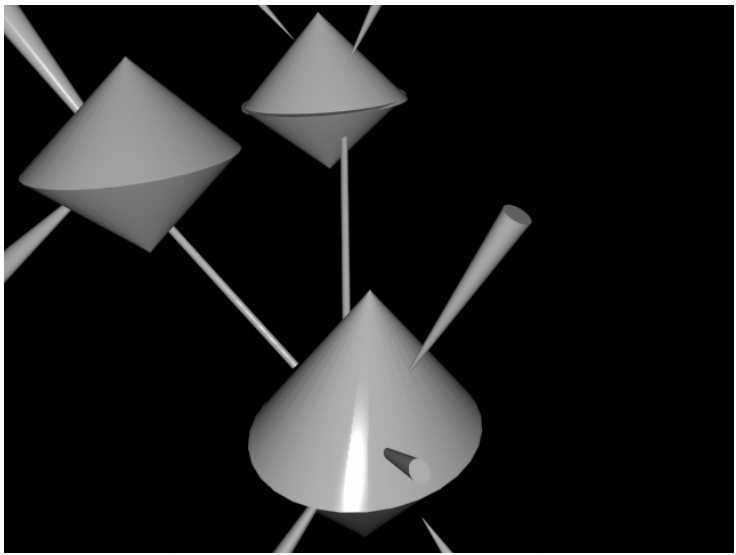}
\end{center}
\vspace{-1.4cm}
\caption{ Exchange diagram involving two jets.\label{exchange}}
\end{figure}
 For massless gravitons with parallel momenta,  the eikonal phases of this diagram and the Newton interaction are supposed to sum up to give a vanishing phase in the S matrix.  It is clear that for generic choice of the Hamiltonian $H(t)$, they will not.  Boost invariance is thus a constraint on the choice of the polynomial in the Hamiltonians of section 1.  It is far from clear to us how strong a constraint it is.  For the case of asymptotically Anti-de Sitter spaces, we know that it is sufficient to tune a few parameters in order to restore the conformal group in the asymptotic limit\footnote{We are implicitly assuming a tensor network regularization of the boundary CFT.}.  It is unlikely that the same will be true here.  Indeed, those cases where we can plausibly recover a quantum gravity scattering operator by taking a limit of a family of conformal field theories, are very rare in the space of all CFTs.  We therefore expect that the constraints of Lorentz invariance are very strong, much stronger than the requirement that a finite system have a limit described by conformal field theory.  Yet another indication of this comes from perturbative string theory, where the naive setup of the perturbation expansion seems to indicate that there are many more string models of quantum gravity than actually exist.  It's only because these models are perturbations of exactly soluble models that we can sort out which models really make sense.  Indeed it's obvious, though perhaps not widely appreciated, that most perturbative string models that have unitary, Lorentz invariant, analytic S matrices to all orders in perturbation theory, but only minimal four dimension SUSY, do not define true models of quantum gravity. 
 
 One disappointing feature of our calculation of the Newton interaction is that it does not seem to come out attractive for arbitrary Hamiltonian in the class we have studied.  We've shown that it comes from the effective Hamiltonian
 \beq H_{eff} (z) = (1 - \Pi) [H_0 + V \Pi (z - H_0)^{-1} \Pi V] (1 - \Pi) . \eeq 
 The interaction comes only from the second term and can be written
 \beq \sum_i (1 - \Pi) V | i \rangle  (z - E_i)^{-1} \langle i | V (1 - \Pi) . \eeq  The sum is over states in the ortho-complement of the constrained subspace.  The large $t$ limit is dominated by $z$ values that are within $1/t$ of the eigenvalues of $H_0$ in the ortho-complement.   In principle, the values of $z$ are determined by finding the the eigenvalues of the $z$ dependent Hamiltonian $H_{eff} (z)$ and then solving the equations, $z_i = E_i (z_i)$ .  These values will give poles of matrix elements of the exact resolvent between states in the constrained subspace.  Since the entire system is finite dimensional, the $E_i (z)$ are values of a multi-sheeted analytic function of $z$ at copies of the real axis on different sheets.  That function also has isolated poles 
 at eigenvalues of $H_0$ on the ortho-complement of the constrained subspace.  
 
 The condition that the expectation value of the interaction term in $H_{eff}$ is negative in typical states in the constrained subspace, seems like a complicated constraint on the parameters in our underlying Hamiltonian.  We had hoped that it would follow from quite general principles, but we do not see our way to such a claim at the moment.
 
 \item A quite satisfactory result of the calculation of Newton's interaction that we have presented is the way in which the notion of energy as a count of the number of constraints appears.   We originally motivated this by referencing black hole entropy formulae, and derived the limiting number of constraints $N_C/T$ as an asymptotic conservation law in all of the models of Section 1.  Here we see the projection on the constrained subspace supplying the factors of energy in Newton's law.  
 \end{itemize} 
 
 To summarize, we've presented a class of explicit, finite quantum mechanical models, all of which have an emergent "space-time interpretation" that manifestly satisfies unitarity, causality, and invariance of the scattering matrix under the subgroup of the Poincare group that preserves a family of time-like trajectories at relative rest.  All of these models exhibit a large distance Newtonian contribution to the scattering matrix of two "localized objects", where the term in quotes is defined in terms of constrained subspaces of the Hilbert space.  The scattering matrix defined by these models has resonances corresponding to long lived metastable states, characterized by an energy, entropy and spatial size that satisfy the parametric relations expected for black holes.  The models are also fast scramblers, in agreement with the properties of black hole quasi-normal modes and have a natural time scale for equilibration that is the Schwarzschild radius.  That same time scale is crucial to the correct scaling of Newton's Law, which can be viewed as arising from excitation of virtual degrees of freedom on the boundary of a causal diamond exactly containing two localized excitations.

 \vskip.3in
\begin{center}
{\bf Acknowledgments }\\
 The work of W.Fischler is supported by the National Science Foundation under Grant Number PHY-1914679.  The work of T. Banks is partially supported by the U.S. Dept. of Energy under grant DE-SC0010008.
\end{center}

\end{document}